\documentclass[%
 reprint,
superscriptaddress,
twocolumn,
nofootinbib,
tabularx,
 amsmath,amssymb
 aps,
]{revtex4-1}

\usepackage{graphicx}
\usepackage{bm}
\usepackage{hyperref}

\newcommand{\RNCaps}[1]
    {\MakeUppercase{\romannumeral #1}}

\begin{document}

\hfill{\footnotesize USTC-ICTS/PCFT-21-46}

\title{Gravitational Waves from Fully General Relativistic Oscillon Preheating}

\author{Xiao-Xiao Kou}
\email{kxx1998@mail.ustc.edu.cn}
\affiliation{
Interdisciplinary Center for Theoretical Study, University of Science and Technology of China, Hefei, Anhui 230026, China
}
\affiliation{Peng Huanwu Center for Fundamental Theory, Hefei, Anhui 230026, China}

\author{James B. Mertens}
\email{jmertens@wustl.edu}
\affiliation{Department of Physics and McDonnell Center for the Space Sciences,
Washington University, St. Louis, MO 63130, USA}

\author{Chi Tian}
\email{chit@wustl.edu}
\affiliation{School of Physics and Optoelectronics Engineering, Anhui University, Hefei, Anhui 230601, China}
\affiliation{Department of Physics and McDonnell Center for the Space Sciences,
Washington University, St. Louis, MO 63130, USA}

\author{Shuang-Yong Zhou}
\email{zhoushy@ustc.edu.cn}
\affiliation{
Interdisciplinary Center for Theoretical Study, University of Science and Technology of China, Hefei, Anhui 230026, China
}
\affiliation{Peng Huanwu Center for Fundamental Theory, Hefei, Anhui 230026, China}

\date{\today}

\begin{abstract}
As long-lived quasi-solitons from the fragmentation of a scalar condensate, oscillons may dominate the preheating era after inflation. During this period, stochastic gravitational waves can also be generated. We quantify the gravitational wave production in this period with simulations accounting for full general relativity to capture all possible non-perturbative effects. We compute the gravitational wave spectra across a range of choices of the oscillon preheating models and compare our results to a conventional perturbative approach on an FLRW background. We clarify the gauge ambiguities in computing induced gravitational waves from scenarios where dense non-perturbative objects such as oscillons are being formed. In particular, we find that the synchronous gauge tends to contain large artificial enhancements in the gravitational wave spectrum due to gauge modes if gravity plays an important role in the formation of the oscillons, while other gauge choices, such as the radiation gauge or a suitably chosen ``1+log'' gauge, can efficiently reduce the contributions of gauge modes. The full general relativistic simulations indicate that gravitational wave spectra obtained from the perturbative approach on the FLRW background are fairly accurate, except when oscillon formation induces strong gravitational effects, for which case there can be an order unity enhancement.
\end{abstract}

\maketitle

\section{Introduction}

During the preheating period after inflation, when the inflaton field still oscillates near the minimum of its potential, strong parametric resonance can trigger 
the exponential growth of fluctuations and quickly push the universe into a highly nonlinear state. In such a phase, fragmentation of the inflaton condensate can generate copious non-perturbative structures, including oscillons \cite{Bogolyubsky:1976pw,Gleiser:1993pt,Honda:2001xg,Amin:2011hj,Amin:2010dc,McDonald:2001iv,Broadhead:2005hn, Copeland:2014qra, Xie:2021glp}. For a large class of inflationary models, oscillons may account for the majority of the energy density during the preheating phase \cite{Amin:2011hj}, which is referred to as the oscillon preheating scenario.

During the oscillon preheating phase, ample relics, such as primordial black holes or stochastic gravitational waves (GWs) can be produced \cite{Zhou:2013tsa, Amin:2018xfe, Liu:2017hua,Antusch:2017flz,Antusch:2017vga,Antusch:2016con,Zhou:2015yfa, Sang:2019ndv,Cotner:2018vug,Cotner:2019ykd, Kou:2019bbc, Nazari:2020fmk,Hiramatsu:2020obh}. While the potential for observable signatures of these relics makes the studies of such phenomena extremely fascinating, the high degree of non-linearity during this phase makes analytical studies very challenging. Thus, numerical approaches that perturbatively solve the Einstein field equations around an FLRW background have been developed and used to quantify the generation of oscillons and their relics \cite{Gleiser:1993pt,Copeland:1995fq,Honda:2001xg,Copeland:2002ku,Farhi:2005rz,Gleiser:2006te,Graham:2006xs,Hindmarsh:2006ur,Fodor:2006zs,Saffin:2006yk,Graham:2006vy,Farhi:2007wj,Hindmarsh:2007jb, Amin:2011hj, Zhou:2013tsa, Amin:2018xfe, Liu:2017hua,Antusch:2017flz,Antusch:2017vga,Antusch:2016con,Zhou:2015yfa, Sang:2019ndv,Cotner:2018vug,Cotner:2019ykd, Kou:2019bbc, Nazari:2020fmk,Hiramatsu:2020obh}. More recently, non-perturbative approaches using techniques from numerical relativity have been applied to preheating scenarios in order to capture all possible nonlinear physics \cite{Bastero-Gil:2007lsx,Bastero-Gil:2010tpb,Giblin:2019nuv,Kou:2019bbc,Nazari:2020fmk}. It has been shown that solving the full Einstein equations will result in an enhancement in the oscillon abundance in models where oscillons are formed with significant help from gravitational attraction, and can even lead to the formation of primordial black holes \cite{Kou:2019bbc,Nazari:2020fmk}.

Recently, an increasing amount of attention has been devoted towards better understanding ambiguities in computing the GW power spectra in different gauges. Those ambiguities originate from the fact that the energy content of GWs is, when working beyond linear order, gauge dependent. This problem was first pointed out in \cite{Matarrese:1997ay}, and has more recently been discussed in \cite{Hwang:2017oxa}, followed by extensive studies \cite{Domenech:2017ems,Gong:2019mui,Tomikawa:2019tvi,DeLuca:2019ufz,Inomata:2019yww,Yuan:2019fwv,Lu:2020diy,Ali:2020sfw,Domenech:2020xin,Gurian:2021rfv}. However, the discussions so far focus on tensor modes induced by second order scalar perturbations around a homogeneous cosmological background. It remains an open question how gauge choice can effect the evaluation of the GW spectrum in the early universe where large inhomogeneities and non-linearities are present in the form of localized and dense structures such as oscillons. 

\begin{figure*}[ht]
\centering
\includegraphics[width=1.0\textwidth]{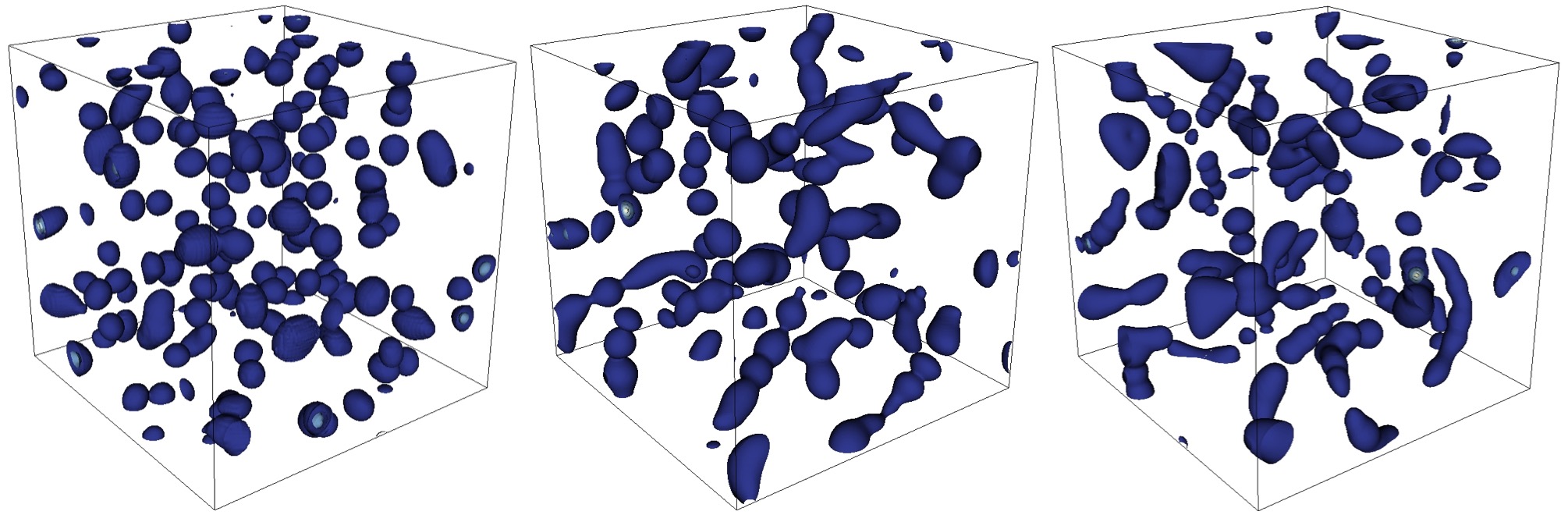}
\caption{From left to right: oscillon formation in simulations with Model \RNCaps{1}, Model \RNCaps{2} and Model \RNCaps{3} respectively, which correspond to different choices of potential parameters $\alpha$ and $\beta$ (see Sec.~\ref{sec:GWs}). Contours of a constant field value are shown in a spatial volume at time $t=187m^{-1}$. Simulations were run using the ``1+log'' gauge choice with $\eta=0.01$.}
\label{fig:fig1}
\end{figure*}

In this paper, we analyze the GWs generated during oscillon preheating with numerical simulations utilizing the full power of numerical relativity. We will see that a perturbative approach on an FLRW background turns out to be a relatively good approximation for the class of models we consider, except when gravitational effects strongly impact oscillon formation, which may lead to enhancements up to about a factor of a few. We will discuss the gauge ambiguities in extracting the GW spectra from the simulations with full general relativity (GR), which is important in the models when oscillons are formed in strong gravity. Our results suggest that the ``1+log'' or the radiation gauge is a more appropriate choice than synchronous gauge when computing GWs beyond linear order, as large artificial enhancements may arise due to gauge modes in synchronous gauge.

The paper is organized as follows. In Section \ref{sec:model}, we will introduce the oscillon preheating model we consider and the methodology we use to numerically solve the system and compute the GWs generated during oscillon preheating. In Section \ref{sec:gauge}, we introduce the gauge conditions we employ with our numerical GR simulations. The main results and discussions are presented in Section \ref{sec:GWs}, where we show the spectra of the GWs computed in different gauges and compare them with the ones calculated with the perturbative method on an FLRW background. We present a semi-quantitative analysis to explain such discrepancies between different gauges in Section \ref{sec:ana}. We summarize and conclude in Section \ref{sec:sum}.

\section{Oscillon Preheating and Models}
\label{sec:model}
As localized quasi-stationary field configurations, oscillons can emerge due to field dynamics in models containing a scalar field with a shallow potential. We will focus on a representative class of minimally coupled models given by the action
\begin{align}
\label{eq:action00}
S=\int \mathrm d^4 x \sqrt{-g} \left( \frac{M_{\rm pl}^2}{2} R - \frac12 \partial_\mu\phi\partial^\mu \phi - V(\phi) \right) ,
\end{align}
with a class of potentials motivated by inflationary model constructions in string/M theory \cite{Silverstein:2008sg, McAllister:2008hb, Dong:2010in},
\begin{align}
\label{eq:V}
V(\phi) = \frac{m^2 M^2}{ 2\alpha} \left[\left(1+\frac{\phi^2}{M^2}\right)^{\alpha} - 1\right].
\end{align}
To facilitate our analysis of the parameter space, we parametrize the potential with the exponent $\alpha$ and the ratio $\beta \equiv M_{\rm pl} / M$, where $M_{\rm pl}$ is the reduced Planck mass. The range of parameters $0<\alpha < 1$ and $0 < \beta < 100$ is considered. Within this range, the potential is flatter than the quadratic mass term at large $\phi$ and approximates to the quadratic term at smaller $\phi$. This type of potential supports quasi-stable oscillon configurations, as the interacting potential, $V(\phi)-\frac12 m^2 \phi^2$, is negative and thus attractive. In other words, ``particles'' under the influence of this kind of potential prefer to condense to localized objects rather than propagate to infinity, which allows long-lived oscillons to form from quite generic initial conditions. A visualization of oscillon formation during the preheating period is shown in Fig.~\ref{fig:fig1}.

The value of scalar field mass $m$ can be estimated by matching to the power spectrum of curvature perturbations from most recent CMB observations  \cite{Aghanim:2018eyx} through the relation
\begin{align}
\label{eq:As}
  A_s = \frac{(4\alpha {\cal N})^{1+\alpha}}{96\pi^2 \alpha^3} \left(\frac{m}{M_{\rm pl}}\right)^2 \!\!  \left(\frac{M}{M_{\rm pl}}\right)^{2-2\alpha} \!\! \simeq 2.1 \! \times\! 10^{-9} ,
\end{align}
where ${\cal N} \simeq 50$ is chosen to be the fiducial number of e-foldings before inflation ends. 

Let us first review how to extract the stochastic GW power spectrum in an FLRW background. Oscillations of the inflaton field obey the Klein-Gordon equation. When coupled to a homogeneous and isotropic cosmological background, the Klein-Gordon equation can be simplified to
\begin{align}
\label{eq:KG}
\ddot{\phi} + 3 H \dot{\phi} - \frac{\nabla^2 \phi}{a^2} + \frac{{\rm d}V(\phi)}{\rm{d}\phi} = 0, 
\end{align}
where $H$ is the Hubble parameter and $a$ is the scale factor. 

The GWs generated around the FLRW background can be quantified by analysing the transverse-traceless part of the metric perturbations, the tensor perturbation $h_{ij}^{TT}$, sourced by matter fields, that is, by solving the linearized tensor perturbation equations
\begin{align}
\label{eq:h1}
\ddot{h}_{i j}^{TT} &+ 3 H \dot{h}_{i j}^{TT}-\frac{1}{a^{2}} \nabla^{2} h_{i j}^{TT}=16 \pi G S_{i j}^{TT}, \\
\label{eq:S1}
S_{i j}^{TT} &= \frac{1}{a^{2}}\left(\partial_{i} \phi \partial_{j} \phi\right)^{TT} .
\end{align}
Numerically calculating $S_{i j}^{TT}(t,\mathbf{x})$ at every step in the simulation is unnecessarily expensive. Rather, due to the linearity of this equation, one can simply evolve $h_{i j}$, which satisfies 
\begin{align}
\ddot{h}_{i j} +3 H \dot{h}_{i j}-\frac{1}{a^{2}} \nabla^{2} h_{i j} =16 \pi G S_{i j}    
\end{align}
and, when needed, project to the physical transverse-traceless part $h_{i j}^{TT}(t,\mathbf{x})$ with:
\begin{align}
h_{i j}^{T T}(t, \mathbf{x})=\int \frac{\mathrm{d}^{3} k}{(2 \pi)^{3}} e^{i \mathbf{k} \cdot \mathbf{x}} \Lambda_{i j, l m}(\hat{\mathbf{k}}) h_{l m}(t, \mathbf{k})
\end{align}
where $\Lambda_{i j, l m}(\hat{\mathbf{k}})$ is a projection operator defined as
\begin{align}
\label{eq:lambda}
\Lambda_{i j, l m}(\hat{\mathbf{k}}) \equiv P_{i l}(\hat{\mathbf{k}}) P_{j m}(\hat{\mathbf{k}})-\frac{1}{2} P_{i j}(\hat{\mathbf{k}}) P_{l m}(\hat{\mathbf{k}})
\end{align}
with $\hat{\mathbf{k}}=\mathbf{k}/k$ and $P_{i j}(\hat{\mathbf{k}})=\delta_{i j}-\hat{k}_{i} \hat{k}_{j}$.
To guarantee tracelessness and transversality in the projection, we use the projector based on nearest-neighbour spatial derivatives suggested by \cite{Figueroa:2011ye} after trying different projectors (see Appendix \ref{sec:proj} for details of tests for different projectors).
The energy density of GWs can be extracted by ensemble averaging of several wavelengths 
\begin{align}
    \label{eq:GWs_frw}
    \rho_{GW} &=\frac{1}{32 \pi G}\left\langle\dot{h}_{ij}^{TT}\dot{ h}^{ijTT}\right\rangle.
\end{align}
The GW energy density spectrum per logarithmic interval is defined as
\begin{align}
\label{eq:rhogw}
\rho_{\mathrm{GW}}(t) &\equiv \int \frac{d \rho_{\mathrm{GW}}}{d \log k} d \log k  , \\
\label{eq:gwspectrum}
\frac{d \rho_{\mathrm{GW}}}{d \log k} &= \frac{k^{3}}{(4 \pi)^{3} G V} \int \frac{d \Omega_{k}}{4 \pi} \dot{h}_{i j}^{TT}(\mathbf{k}, t) \dot{h}_{i j}^{* TT}(\mathbf{k}, t)  ,
\end{align}
where $\Omega_k$ represents the solid angle in $\boldsymbol{k}$-space, and the volume $V=L^3$ in the lattice implementation. It is customary to normalize the GW energy density spectrum with the critical density $\rho_{c} \equiv 3 H_0^{2} / 8 \pi G$, leading to
\begin{align}
\label{eq:normalgws}
\Omega_{\mathrm{GW}}(k, t) \equiv \frac{1}{\rho_{c}} \frac{d \rho_{\mathrm{GW}}}{d \log k}
\end{align}

This is the conventional perturbative approach on an FLRW background, which involves relatively simple differential equations and where the non-linear dynamics of oscillons and their GW generation can be well resolved with fairly low computational costs. However, in this approach, gravitational backreaction is neglected, which {\it a priori} may not be a valid approximation for simulations when gravitational effects are strong during the oscillon formation. Indeed, it was found in \cite{Kou:2019bbc} that, with the full GR approach, the abundance of oscillons in preheating is enhanced for those circumstances.

We shall investigate whether solving full Einstein equations might introduce corrections to the relic GW spectra. To this end, we perform the 3+1 decomposition of the Einstein equations, which decomposes the metric into the form
\begin{align}
\label{eq:metric}
g_{\mu\nu} = 
\begin{bmatrix}
-N^2 + N_kN^k & N_j \\
N_i       & \gamma_{ij}   
\end{bmatrix},
\end{align}
where $N$ and $N^i$ are the lapse and shift respectively and $\gamma_{ij}$ is the 3D metric on the spatial hypersurfaces foliated by $N$ and $N^i$. The extrinsic curvature tensor $K_{ij}$ describing the embedding of the spatial hypersurfaces is defined as the Lie derivative of $\gamma_{ij}$ with respect to the normal vector of the hypersurfaces, namely 
\begin{align}
    K_{ij} = \mathcal{L}_{n}\gamma_{ij}.
\end{align}
We also cast the Klein-Gordon equation coupled to GR into the hyperbolic form 
\begin{align}
\label{eq:5}
\dot{\phi} &= N^i \partial_i\phi -N \Pi
\\
  \dot{\Pi}   &=    N^k\partial_k \Pi - N \gamma^{ij} \partial_i  \psi_j + N \gamma^{ij} \Gamma^{k}_{ij}\psi_k  +  N K \Pi \nonumber \\
                &-   \gamma^{ij} \psi_i \partial_j N  +  N {\rm d}V(\phi)/{\rm d}\phi \\
 \dot{\psi}_i &= N^j\partial_j \psi_i + \psi_j \partial_i N^j -N \partial_i \Pi -\Pi \partial_i N,
\end{align}
where $\psi_i$ is the auxiliary field $\psi_i \equiv \partial_i \phi$, $K$ is the trace of the extrinsic curvature and $\Pi$ is the canonical momentum of the scalar field 
\begin{align}
\label{eq:8}
\Pi \equiv - \left(\frac{1}{N}\dot{\phi} - \frac{1}{N} N^k\partial_k\phi \right) .
\end{align}
We will from now on refer to this scheme as the \textit{GR scheme}, and the scheme relying on Eq.~\eqref{eq:KG} as the \textit{FLRW scheme}.  

To compute the energy of GWs in full GR and compare with the FLRW scheme, we need a background expanding metric. A nature choice is to average the spatial hypersurface to get the background, then subtract the expanding background from the full metric \cite{Isaacson:1967zz,Isaacson:1968zza,BasteroGil:2010nm,Ota:2021fdv}. This allows us to define 
\begin{align}
\label{gwsgr}
\rho_{GW} &=\frac{1}{32 \pi G}\left\langle{\hat{\gamma}}_{ij;0}^{TT}{\hat{ \gamma}}^{ijTT}_{;0}\right\rangle,
\end{align}
where ${}_{;0}$ means covariant time derivative compatible with background metric.
And $\hat{\gamma}_{ij}$ is defined as
\begin{equation}
    \hat{\gamma}_{ij} \equiv \chi^{-2}\Tilde{\gamma}_{ij} - a^2(t)\delta_{ij},
    \label{eqn:pertmetric}
\end{equation}
where $\chi$ is the conformal factor and $\tilde{\gamma}_{ij}$ is the conformal metric defined by decomposing the spatial metric $\gamma_{ij}$ as $\tilde{\gamma}_{ij} \equiv \chi^2 \gamma_{ij}$, with the constraint ${\rm det}\tilde{\gamma}_{ij}=1$.  When the metric is conformally flat ($\tilde{\gamma}_{ij} = \delta_{ij}$) and $\chi$ has no spatial dependence, $\chi^{-1}$ is equivalent to the scale factor $a$. However, the inhomogeneous distribution of the scalar field will introduce small spatial dependence to $\chi$ through the Einstein equations. Its average still has the time evolution that approximately corresponds to the scale factor in the FLRW universe, which allows us to use $\langle\chi^{-1} \rangle$ to mimic the scale factor in Eq.~\eqref{eqn:pertmetric}. Note that $\hat{\gamma}_{ij;0}^{TT}$ essentially reduces to the $2K_{a b}^{\mathrm{TT}}$ of \cite{Ota:2021fdv} when the fluctuations in the field $\chi$ are very small, and these small fluctuations can be initially fixed by solving the constraints equations. Also note that non-perturbatively defining a GW energy that is gauge independent and can be implemented in a grid simulation scheme is highly nontrivial. Indeed, the $\rho_{GW}$ defined above, as we will see later, is gauge dependent. As we will explain later, we will need to choose gauge conditions judiciously so as to minimize the gauge effects from the redundant gauge modes. In particular, despite the background being in synchronous gauge, we will see later that the synchronous gauge may not be a good gauge choice for the perturbations on top of the background.

To solve the metric components together with the scalar field they couple to, we employ the grid-based numerical relativity code \textsc{CosmoGRaPH} \cite{Mertens:2015ttp}, which makes use of the popular Baumgarte--Shapiro--Shibata--Nakamura (BSSN) formalism \cite{Nakamura:1987zz,Shibata:1995we, Baumgarte:1998te} and integrates the Adaptive Mesh Refinement (AMR) framework into its spatial grid scheme. However, for simulations performed in this work, only uniform grid configuration is employed. 

 Different from the FLRW scheme, solving the equations in the GR scheme requires  obtaining exact solutions to the constraint equations in the initial condition setup. We will seek solutions that mostly resemble the FLRW metric to facilitate our comparisons between the two schemes. To this end, after splitting the extrinsic curvature into a traceless part and a trace part $K_{ij} \equiv A_{ij}+\frac13 \gamma_{ij} K$, we choose the initial conditions such that $\tilde{\gamma}_{ij} = \delta_{ij}$, $A_{ij} = 0$. Since $K=-3H$, the initial homogeneous $K$ is set such that the corresponding Hubble parameter satisfies the Friedmann equation. The initial configuration of the scalar field is chosen to have a standard spectrum of the initial vacuum fluctuations for $\phi=M$ and $\dot{\phi}=0$, for which the momentum constraint is trivialized, and the Hamitonian constraint equation can be written as
 \begin{equation}
     \label{eq:H}
     \nabla^2 \Psi + \pi \left(\partial_i \phi \partial_i \phi\right) \Psi + \left(\pi V(\phi)  - K^2 / 12  \right) \Psi^5 =0,  
 \end{equation}
 where $\Psi \equiv \chi^{-1/2}$.
 We then solve this non-linear Hamiltonian constraint equation and get the conformal factor $\chi$ by employing the multigrid constraint solver integrated within \textsc{CosmoGRaPH}. We use a periodic box with a spatial size $L = 50\, m^{-1}$. To ensure numerical convergence of the elliptic constraint solver, a cut-off at wavenumber $k=12\,(2\pi / L)$ is implemented.

\begin{figure*}
\centering
\includegraphics[scale=0.52]{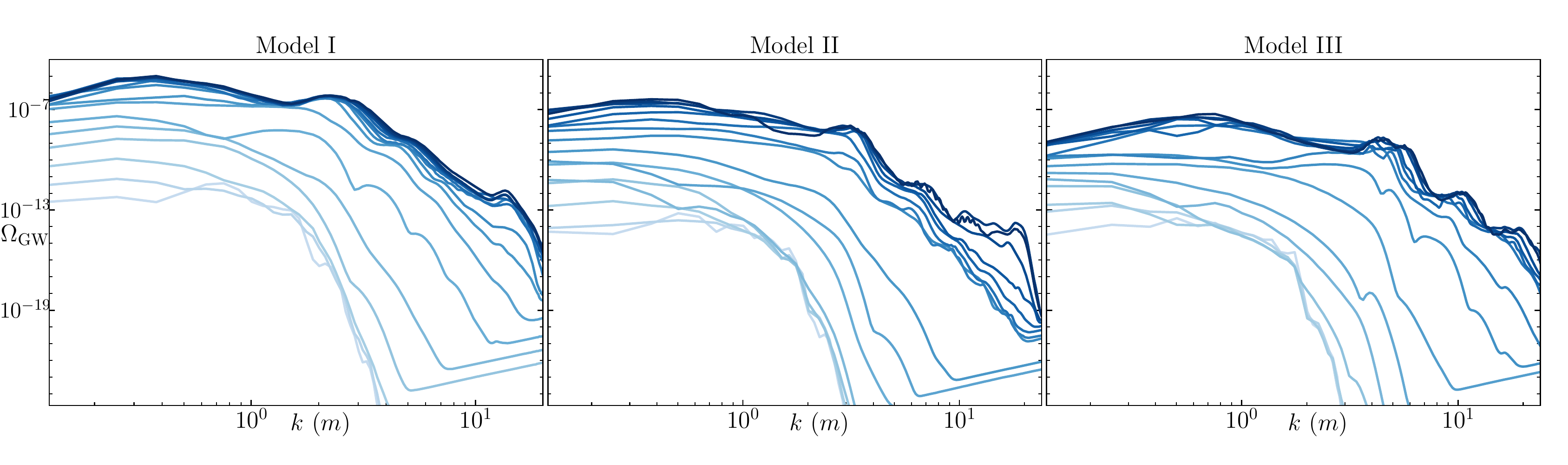}
\caption{Time evolution of GW power spectra in the FLRW scheme, with darker colours representing later times. Model \RNCaps{1} to \RNCaps{3} are presented from left to right.}
\label{fig:gw_grow}
\end{figure*}

\section{Gauge choices}
\label{sec:gauge}
Even though the perturbative definition of GW energy Eq.~\eqref{eq:GWs_frw} is a gauge invariant quantity to linear order (when the box size is larger than the wavelengths of GWs, see \cite{misner1973gravitation}), it becomes gauge dependent when higher order perturbations are included. This gauge dependence is by construction absent from the FLRW scheme, as the Klein-Gordon equation is only coupled to the background cosmology (Eq.~\eqref{eq:KG}). 
We will consider multiple gauge choices in the GR scheme to assess how large the gauge dependence is for our simulations. By comparing different gauges, we look for gauge choices that are insensitive to the gauge modes from higher order perturbations, which will allow us to establish discrepancies between the GR and the FLRW scheme.

 The simplest gauge condition we consider here is the synchronous gauge, or geodesic slicing, which simply sets the lapse $N=1$ and a vanishing $N^i$. Indeed, our background is in the synchronous gauge. Despite its simplicity, the synchronous gauge tends to focus geodesics and create coordinate singularities very easily. In addition, it is well known that the synchronous gauge is not a complete gauge fixing, even to leading order. As we shall see shortly, we find this gauge generally an unsuitable choice for extracting GW spectra from oscillon preheating.

Another widely used set of gauge conditions in numerical relativity is the combination of the ``1+log'' and the ``Gamma-driver'' shift conditions:
\begin{eqnarray}
\label{eq:gaugecondition}
  \partial_t N &=& -2 \eta N \left(K - \langle K \rangle\right) + N^i \partial_i N,  \\
  \partial_t N^i  & = & B^i, \;\;\;\;  \partial_t B^i  =  \frac34 \partial_t \tilde{\Gamma}^i - B^i, 
\end{eqnarray}
where $\eta$ is a constant to be chosen, $B^i$ is only an auxiliary field and $\tilde{\Gamma}^i \! \equiv \! \tilde{\gamma}^{jk} \tilde{\Gamma}_{jk}^i$, with $\tilde{\Gamma}_{jk}^i$ being the Christoffel symbols of $\tilde{\gamma}_{ij}$.
We will refer to this gauge condition as the ``1+log'' gauge for simplicity. This gauge condition has been proven to have a powerful singularity-avoiding property, and thus is widely used in solving spacetime evolutions containing black holes. 

We also examine a less-used gauge choice, called the radiation gauge \cite{Chen:2010wf}: $g^{ij}\Gamma_{ij}^{\rho} = 0$. This gauge condition might look like the conventional harmonic gauge $g^{\mu\nu}\Gamma_{\mu\nu}^{\rho} = 0$ in appearance, but it is rather different in nature. In fact, it can be proven to be equivalent to time evolution of the lapse according to:
\begin{align}
\label{eq:rad}
\partial_t N = -\frac{2\tau N^2}{1+N^2}\left(K - \langle K \rangle\right) ,
\end{align} 
with a vanishing shift $N^i$, where $\tau$ is chosen to be 0.05 in this paper. Note that this gauge condition has the late-time asymptotic solution that $K\rightarrow \langle K\rangle$ and $\partial_t N = 0$, which corresponds to a homogeneous expansion rate.

\section{Quantifying the GW spectra}
\label{sec:GWs}

\begin{figure*}
\centering
\includegraphics[scale=0.103]{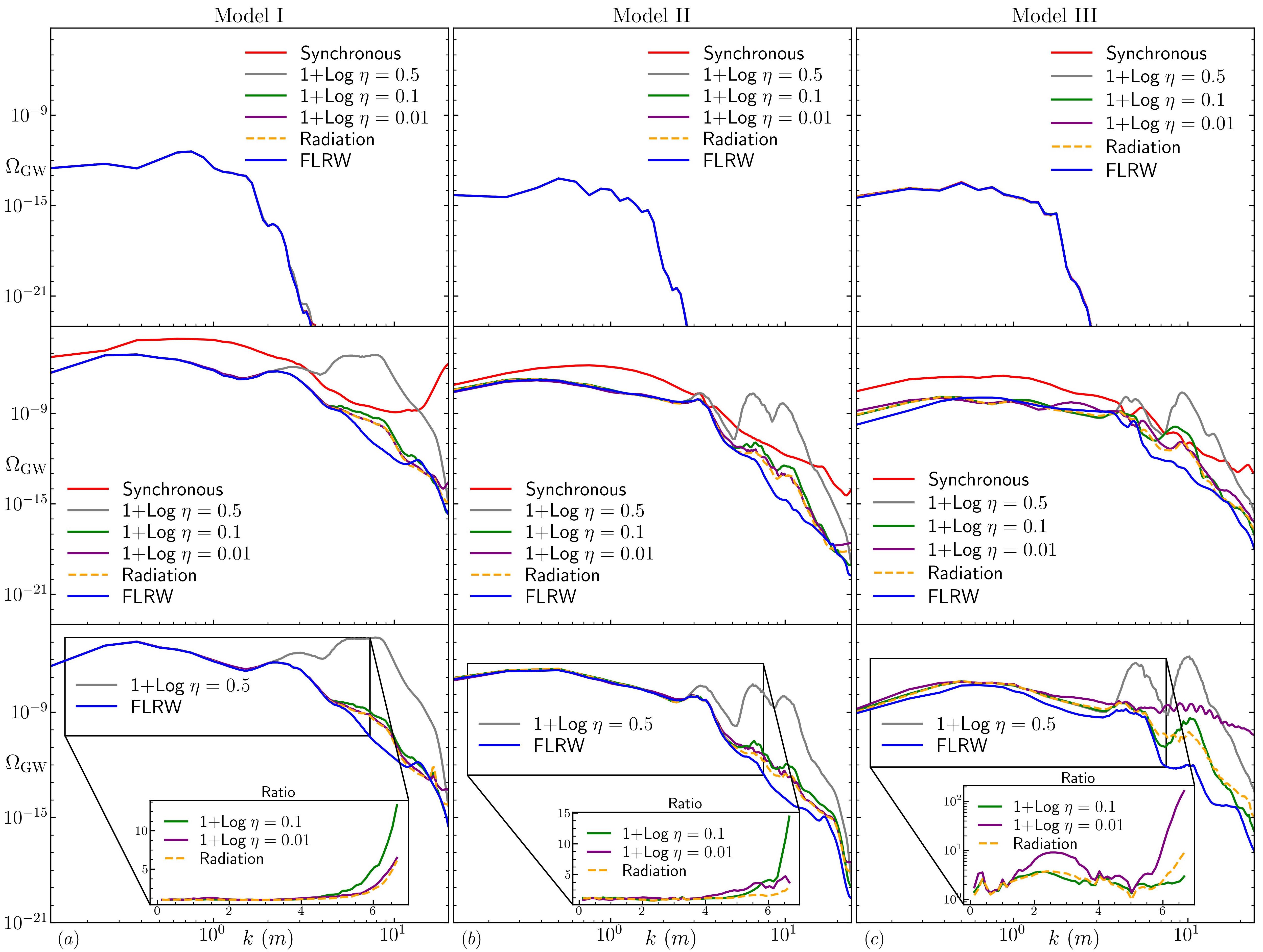}
\caption{Time evolution (from top to bottom) of GWs spectra for $3$ sets of Models with different gauge choices. $(a)$ Model I -- strong resonance: The top panel is taken at $t\approx 6m^{-1}$; the middle panel, at $t\approx 94m^{-1}$, is when the simulation in the synchronous gauge breaks down; the bottom panel, at $t\approx 112m^{-1}$, shows the GW spectra in the final stage of preheating. $(b)$ Model II -- median resonance: The top panel is taken at $t\approx 8m^{-1}$; the middle panel, at $t\approx 156m^{-1}$ is when the simulation in the synchronous gauge breaks down; the bottom panel, at $t\approx 165m^{-1}$, shows the spectrum in the final stage of preheating. $(c)$ Model III -- weak resonance and strong self-gravity: The top panel is taken at $t\approx 15m^{-1}$; the middle panel, at $t\approx 140m^{-1}$ is when the simulation in the synchronous gauge breaks down; the bottom panel, at $t\approx 172m^{-1}$, shows the spectrum in the final stage of preheating. The insets in the bottom panels depict the ratio between the GW spectrum in the corresponding gauge condition and that of the FLRW simulation. For Model I, we run the simulation based on a $192^3$ grid, but for Model II and III, we utilize $256^3$ simulations to capture potential self-gravity effects.}
\label{fig:gws_evol}
\end{figure*}

We select three representative sets of model parameters to show the influence of the gauge choice and possible enhancements of GW power spectra compared with the conventional FLRW based scheme.
\begin{center}
\begin{tabular}{ |l|l| }
\hline
 Model \RNCaps{1} & $\alpha=0.18,\beta=50$ \\ 
  \hline
 Model \RNCaps{2} & $\alpha=0.50,\beta=50$  \\
  \hline
 Model \RNCaps{3} & $\alpha=0.18,\beta=22$ \\ 
\hline
\end{tabular}
\end{center}
Model \RNCaps{1} has very strong parametric resonance and generates a large number of oscillons in the preheating period, Model \RNCaps{2} corresponds to the case with slightly weaker parametric resonance, and Model \RNCaps{3} corresponds to the case with very weak parametric resonance from the scalar self-interaction but with strong self-gravity during the oscillon formation\,\footnote{Some authors may refer to an oscillon with strong self-gravity as an oscillaton.}.  Note that even though the scalar parametric resonance in Model \RNCaps{3} is weak, strong gravitational effects can boost the generations of oscillons (see \cite{Kou:2019bbc} for more detail). 

For all the 3 sets of parameters of the potential, the inflaton condensate can efficiently fragment into oscillons; see Fig.~\ref{fig:fig1} for the illustrations of oscillons for the $3$ models in the GR scheme. During the fragmentation, significant stochastic GW backgrounds are generated, and their power spectra keep growing until the oscillons are stabilized. We compute the power spectra of GWs $\rho_{\rm GW}(k)$ after oscillons form and are stabilized, and the growth of the GWs spectra for all $3$ models is shown in Fig.~\ref{fig:gw_grow}. Note that at the earlier stage during the evolution, the high frequency ends of the spectra have approximated $k$ dependence $\Omega_{\rm GW} \propto k^3$, which result in straight lines in the lower right corner in this logarithmic plot. This is due to the numerical artifact of applying the cutoff at higher $k$ --- the cutoff results in a small and flat spectrum of $h_{ij}^{\rm TT}$ at higher $k$, which translates into the relation $\Omega_{\rm GW} \propto k^3$ approximately according to Eq.~\eqref{eq:gwspectrum}.

We depict the evolution of the GW spectra with different gauges for all $3$ models in Fig.~\ref{fig:gws_evol}.
The GWs from Model \RNCaps{1}, which gives rise to strong parametric resonance and formation of a large number of oscillons in preheating, is plotted in the left column for 3 time instances in Fig.~\ref{fig:gws_evol}. We can see that, for the ``1+log" gauges and the radiation gauge, the results generally coincide with those of the FLRW scheme until the wavelengths are significantly smaller than the box size. The enhancements seen for these small wavelengths are mostly numerical artifacts due to the lack of spatial resolution for these modes (see Appendix \ref{sec:conv} for details of the convergence tests).
These results are consistent with the GW spectra computed in \cite{Zhou:2015yfa}, which uses a pseudo-spectral method and evolves the metric perturbatively on an FLRW background. The multi-peak patterns in the spectra are intimately linked to the fact that oscillons evolve to a spherically symmetric shape after emerging from the inflaton condensate fragmentation \cite{Zhou:2015yfa}. We also note that the choice of the synchronous gauge stands out as an exception, where an order of magnitude of enhancements are observed before the simulation breaks down at $t\approx 95 m^{-1}$ (see the middle panel in the left column in Fig.~\ref{fig:gws_evol}). This suggests that there are huge redundant gauge modes in evaluating the GW production in oscillon preheating with the synchronous gauge, meaning that the synchronous gauge is a rather inappropriate choice for calculating the GW production in full numerical relativity, at least for oscillon preheating. 

For Model \RNCaps{2}, as shown in the middle column in Fig.~\ref{fig:gws_evol}, similarly as in the Model \RNCaps{1}: the GW spectra in ``1+log" gauges and the radiation gauge are also consistent with the GW spectra from the FLRW scheme, within the wavelengths that are well resolved numerically. Synchronous gauge breaks down at around $156\,m^{-1}$, where the GW power spectrum is again significantly enhanced due to large artificial gauge modes.

As suggested by \cite{Kou:2019bbc}, Model \RNCaps{3} is subjected to strong gravitational backreaction, which can only be adequately resolved by the full GR scheme. As shown in the right column in Fig.~\ref{fig:gws_evol}, apart from similar large gauge modes in synchronous gauge (red line in the middle panel), the GW spectra computed in ``1+log'' gauges are also subjected to some gauge dependence for the wavelengths that are well resolved in our simulations. We find that the gauge with $\eta = 0.1$ is a better choice compared to that with $\eta = 0.01$ and $\eta = 0.5$, as it contains relatively fewer gauge modes. The reason for this is that: when $\eta$ approaches 0, the 1+log gauge choice approaches the synchronous gauge, which has been shown to have a rather large gauge redundancy; when $\eta$ is large, on the other hand, it is overdamping the lapse function. An overdamped lapse function will make the high frequency part within $\tilde{\gamma}_{ij}$ less stable compared with the choice of $\eta=0.1$, and these unstable high frequency gauge modes will enter $\hat{\gamma}_{ij}^{TT}$ within its evolution, as can be seen in the high frequency peaks for $\eta = 0.5$ in the last panel of Fig.~\ref{fig:gws_evol}.

Also, the GW spectra from ``1+log'' gauge with $\eta = 0.1$ is consistent with radiation gauge with $\tau = 0.05$ (cf.~Eq.~\eqref{eq:rad}), which suggests that the GW spectra in these gauges might be taken as a good estimate of the ``physical'' spectra in this model. With this interpretation, a small enhancement of a factor of $3$ in the medium wavelengths can identified.

Similar to the gauge ambiguities of the GW energy density we have shown in the full GR simulations, the gauge dependence of GWs in higher order cosmological perturbation theory has been widely discussed in the context of GWs induced by higher order scalar perturbations \cite{Matarrese:1997ay,Hwang:2017oxa,Tomikawa:2019tvi,Gong:2019mui,Lu:2020diy} (see \cite{Domenech:2021ztg} for a recent comprehensive review). In these scenarios,  synchronous gauge is often found to contain more gauge modes and thus is less reliable in evaluating quantities related to GWs. Similar gauge issues are also identified here in the oscillon preheating model we study. We show that when oscillons are in a stably oscillating mode, some gauges do contain fewer gauge ambiguities, and are thus more suitable for performing computations and comparing with observations. On the other hand, as pointed out recently \cite{Inomata:2019yww,Yuan:2019fwv,Domenech:2020xin}, the differences between different gauge choices will eventually decrease and die down when the sources that induce those GWs decay at a later stage and gravity becomes weaker. As our observations are always at much later times, these gauge ambiguities thus may not present any true difficulties in theory, if we can follow the evolution for a sufficiently long time. For our current case, this means the time when oscillons decays, which is technically challenging, as oscillons in our model are very stable. We leave this line of investigation for future work.   

\section{Gauge dependence of GW energy density}
\label{sec:ana}
In Fig.~\ref{fig:gws_evol}, we can observe at least one order of magnitude enhancement of GW power spectra in the synchronous gauge before the simulation breaks down, compared with properly chosen ``1+log'' gauges and radiation gauge. In this section, we will show how such big gauge dependence may arise by performing a semi-quantitative analysis of higher order perturbations in the GWs content.

\begin{figure}
\centering
\includegraphics[scale=0.5]{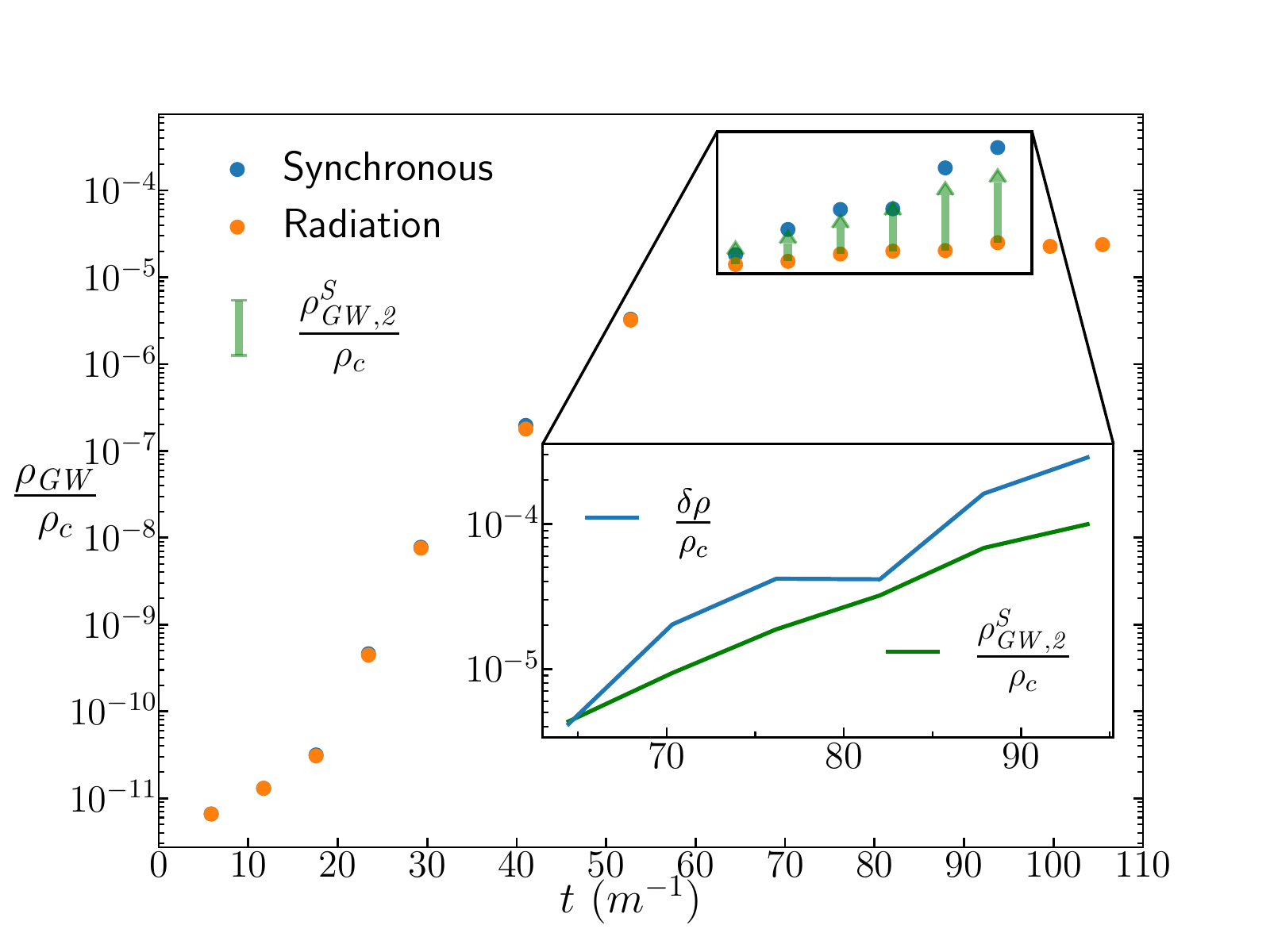}
\caption{Time evolution of GW energy densities for Model \RNCaps{1} with the synchronous and the radiation gauge. The inset depicts the estimate of the second order gauge dependence $\rho_{GW,\rm 2}\sim {T}{h}\delta{h}$ as well as the difference of the GW energy density between the synchronous and the radiation gauge. Both energy densities are extracted from the full GR simulation of Model \RNCaps{1} and normalized to the critical density at the time when the simulation with the synchronous gauge breaks down.}
\label{fig:gauge}
\end{figure}

We want to compare the GW energy density (cf. Eq.~(\ref{eq:GWs_frw})) between the two different gauges, i.e. between two different coordinate systems. Consider small coordinate transformation
$x^{\alpha} \rightarrow \Tilde{x}^{\alpha}=x^{\alpha} + \epsilon\xi^{\alpha}$
which induces gauge transformation
\begin{align}
\label{eq:pert}
\Tilde{h}_{ij} \simeq h_{ij} - \partial_j\xi_i -  \partial_i\xi_j - h_{ik}\partial_j\xi^k - h_{jk}\partial_i\xi^k + O(\epsilon^3) .
\end{align}
where we have neglected terms of order $\epsilon^3$ and terms that, by integrating by part, lead to divergences when evaluating the GW energy density. In the $\tilde x^\alpha$ coordinates, the terms $\partial_j\xi_i +  \partial_i\xi_j$ do not contribute to the GW energy density, thanks to the transverse-traceless projection on $h_{ij}$, which means that the GW spectrum is gauge invariant to leading order. However, in the full GR scheme, higher order terms like $h_{ik}\partial_j\xi^k$ do give rise to nonvanishing contributions to the GW energy density in the $\tilde x^\alpha$ coordinates, which leads to gauge dependence in the GW energy density for higher order perturbations. Because 
the GW energy density is represented schematically as $\rho_{GW} \sim\partial h\partial h$, the contributions from the higher order terms in the GW energy content representing the gauge modes can be estimated by
\begin{align}
\label{eq:est}
\rho_{GW,\rm 2} &\sim \partial h \partial(h\partial\xi) + O(h^4) \nonumber\\
&\sim (\partial^2 h)h\left(\partial\xi\right) \sim T h\delta h
\end{align}
where we have used the leading order equation of motion of $h_{ij}$ to replace $\partial^2 h$ with the matter stress-energy tensor $T_{ij}$, and also approximated $\partial\xi$ with $h-\tilde{h}$. As a simple estimate of the GW energy density from the gauge modes, we will simply approximate $\rho_{GW,\rm g}$ with $\rho^{S}_{GW, \rm 2}=(\sum T)(\sum h) (\sum \delta h)$, where $\sum T$ stands for $\sum_{ij} T_{ij}/6$ and so on. We compare this estimation with the difference in the GW energy density between the synchronous gauge and the radiation gauge in Fig.~\ref{fig:gauge}, where the green arrows representing the second order estimations for gauge modes approximately match the difference between the two gauges. It is also worth noting that Eq.~\eqref{eq:est} implies that the gauge modes decay when the source $T$ is suppressed.

\section{Summary}
\label{sec:sum}
In this paper, we have studied the production of stochastic GWs during the oscillon preheating scenario by numerically solving the full Einstein equations under different gauge conditions. The results show that in the models where gravitational backreation is negligible during oscillon preheating, the GWs extracted from full GR simulations coincide with the results given by the conventional simulations in an FLRW background, except for the spectrum computed in the synchronous gauge. In the synchronous gauge, the computed GW power spectrum is typically 1-2 orders of magnitude higher than that of the FLRW simulation or GR simulation with the ``1+log'' or the radiation gauge, which means that significant redundant gauge effects are present in this gauge, making it an unsuitable gauge condition to be used in oscillon preheating. On the other hand, for models where gravitational backreaction is significant, the stochastic GW power spectra are enhanced by up to a factor of $3$ in the full GR simulations under proper gauge choices.

The GWs generated by the particular single-field inflationary models considered in this paper have a frequency band beyond the current and next generation of GW detectors. However, once the role of the inflaton, which should correctly generate the power spectrum of the observed cosmic microwave background, is released for the scalar field that generate oscillons, an oscillon preheating-like scenario can occur at lower energy scales, and the frequencies of the generalized GW background can be in the bands detectable in the GW detectors. Potential detection of GW signals for such models can be used to test general relativity and probe new physics at high energy scales and in the early universe.

Our study suggests that it is important to choose appropriate gauge conditions to evaluate GW power spectra in full GR preheating simulations, and the synchronous gauge appears to be a very unreliable gauge condition to be used to extract GW spectra in full GR simulations, at least when dense objects such as oscillons are being formed. Incidentally,  \cite{Bastero-Gil:2010tpb} found an order of magnitude enhancement for the stochastic GW spectrum in a hybrid inflation model, using the full GR simulations with the synchronous gauge. Of course, in that model no oscillons or other non-perturbative objects are forming during the preheating. Nevertheless, it would be interesting to revisit it and exclude any possible gauge redundancies, which are left for future work.

\section*{Acknowledgements}

We are grateful to Li-ming Cao, Glenn Starkman and John T. Giblin, Jr for helpful discussions. SYZ acknowledges support from the starting grants from University of Science and Technology of China under grant No.~KY2030000089 and GG2030040375, and is also supported by National Natural Science Foundation of China under grant No.~11947301, 12075233 and 12047502, and supported by the Fundamental Research Funds for the Central Universities under grant No.~WK2030000036.
JBM is partly supported
by a Department of Energy grant DE-SC0017987 to the
particle astrophysics theory group at WUSTL

\appendix

\section{Convergence Tests}
\label{sec:conv}
In this work, we compute GWs spectra in a periodic box with a spatial size $L = 50/m$. For Model I, we run the simulations using a $192^3$ grid, since self-gravity effects are not strong when the parametric resonance is significant, and the resolution of $192^3$ is enough to capture essential physics. For Models II and III, we utilize $256^3$ simulations to capture potential self-gravity effects, as fully general relativistic corrections are often important when the parametric resonance is weak \cite{Kou:2019bbc}. The validity of these choices of resolutions is also examined by the convergence tests.

\begin{figure}[h]
\centering
\includegraphics[scale=0.5]{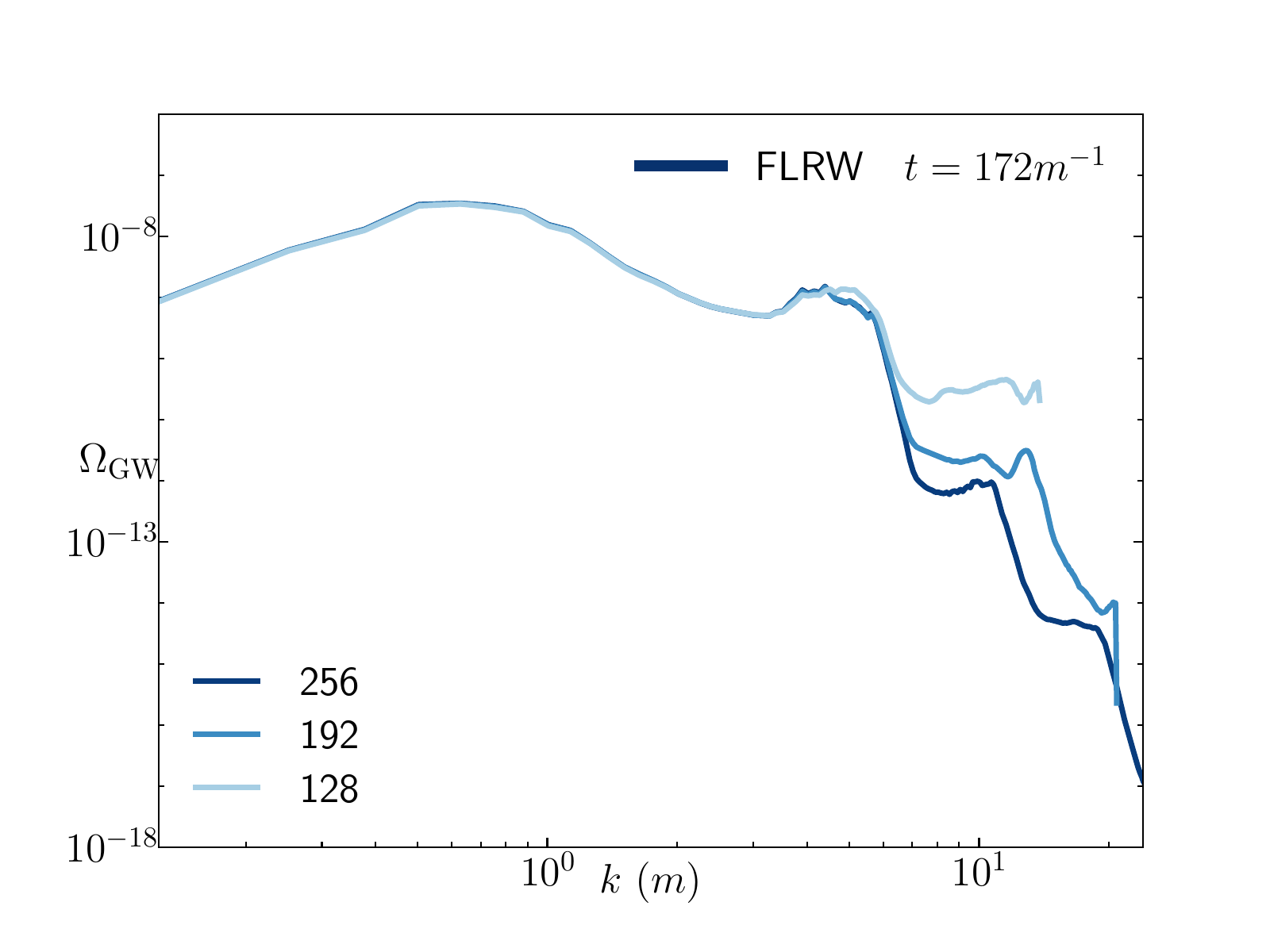}
\caption{GW power spectra in the FLRW scheme under different resolutions.}
\label{fig:flrwtest}
\end{figure}

\begin{figure}[h]
\centering
\includegraphics[scale=0.5]{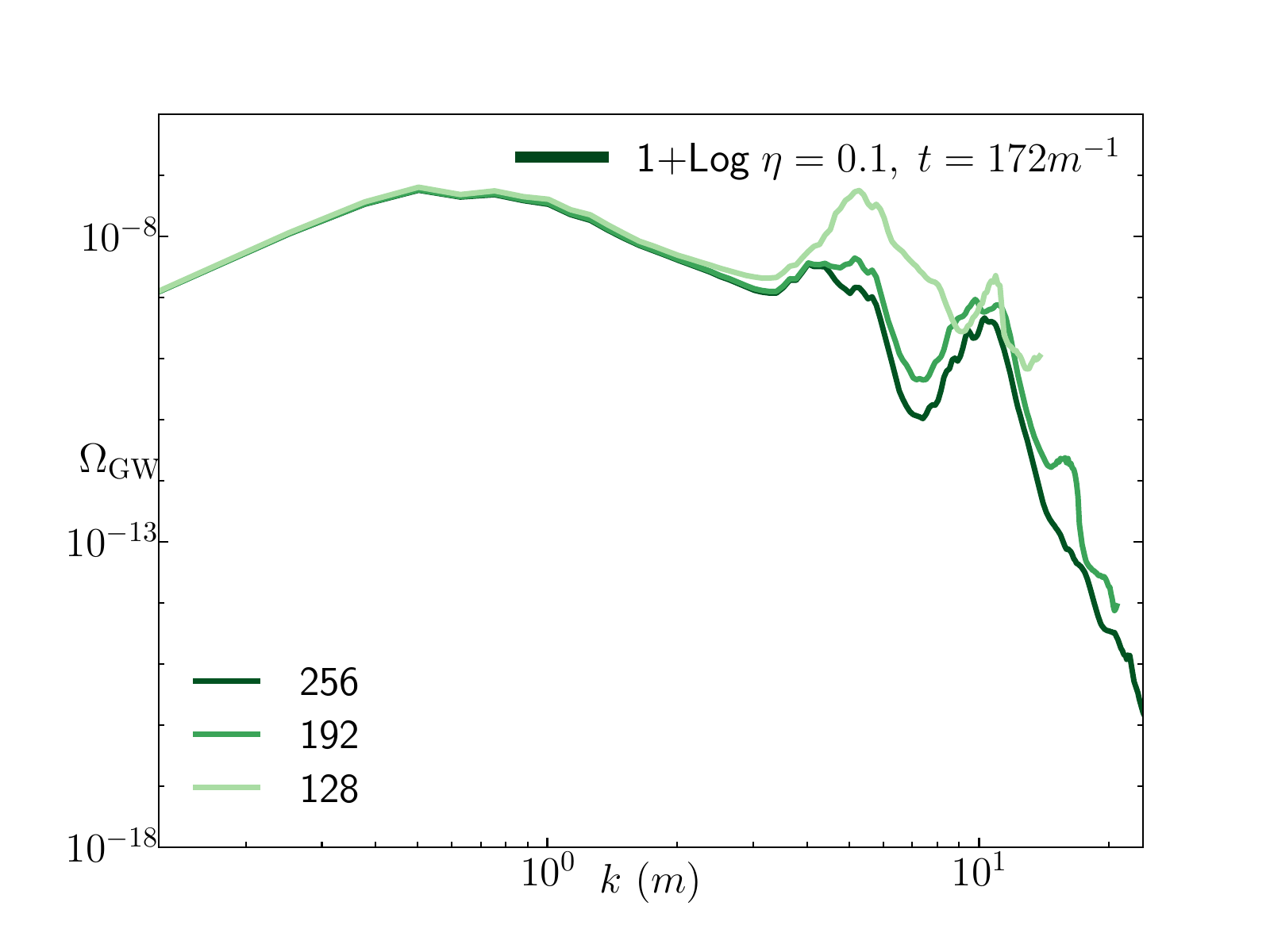}
\caption{GW power spectra in the ``1+log'' gauge with $\eta=0.1$ under different resolutions.}
\label{fig:fractest1}
\end{figure}

\begin{figure}[h]
\centering
\includegraphics[scale=0.5]{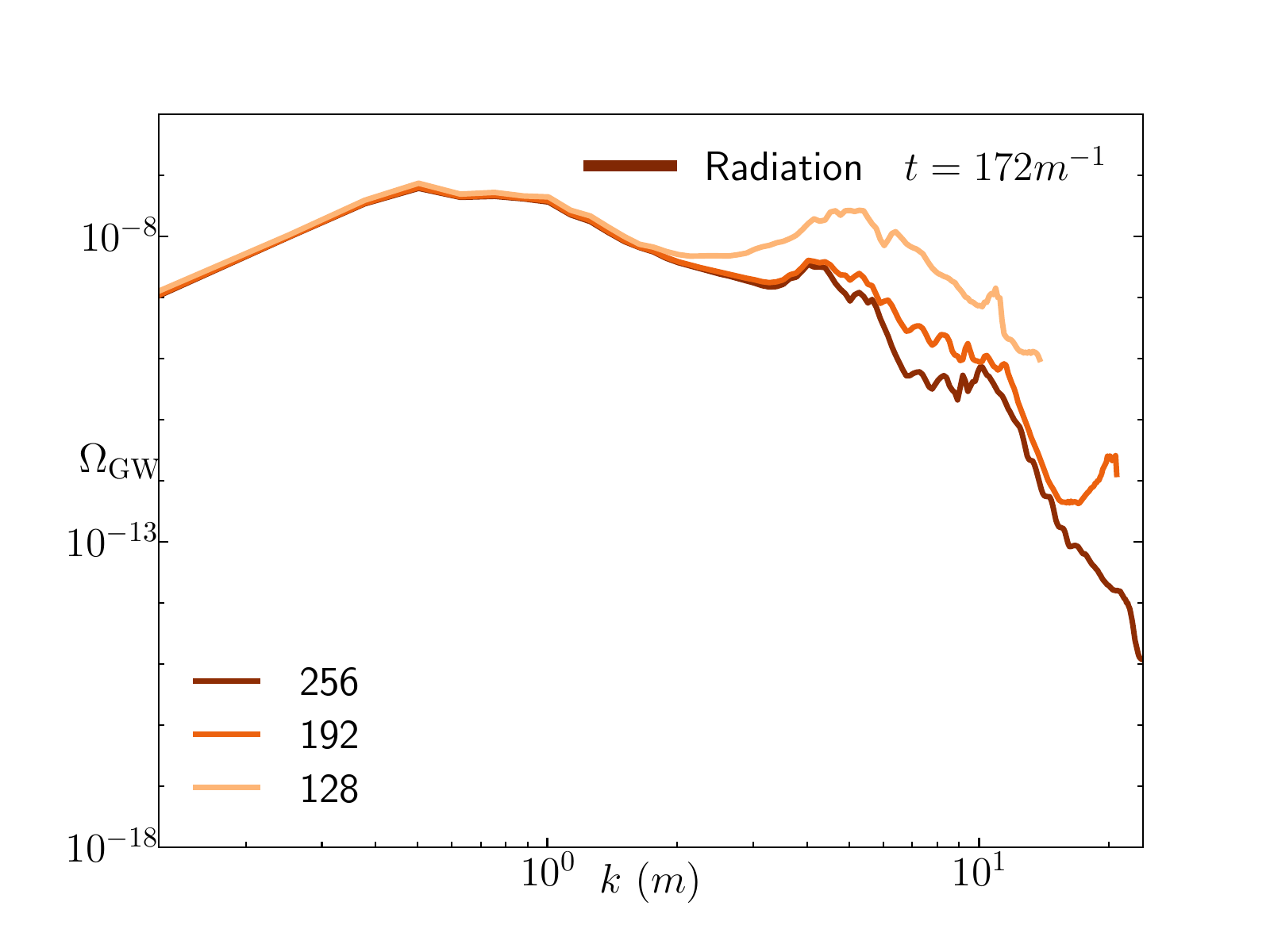}
\caption{GW power spectra in the radiation gauge under different resolutions.}
\label{fig:fractest2}
\end{figure}

\begin{figure}[h] 
\centering
\includegraphics[scale=0.5]{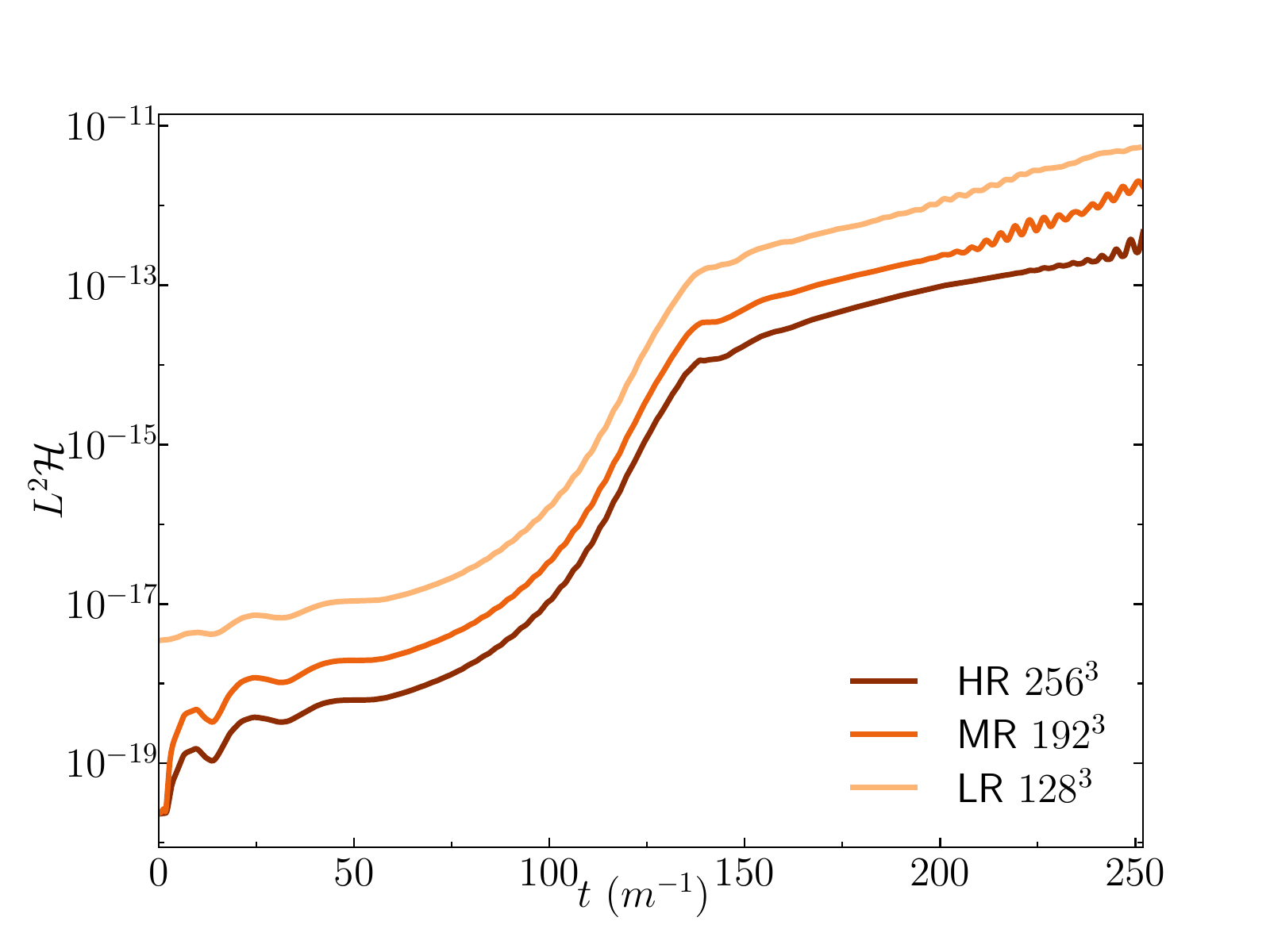} 
\caption{L2 norms of the Hamiltonian constraint under different resolutions in the radiation gauge. The model parameters are chosen as $\alpha=0.18$ and $\beta=22$.}
\label{fig:constraints}
\end{figure}

Here we show the convergence tests for the most demanding case, Model \RNCaps{3}, which is the model that needs the most simulation steps and converges the slowest. The convergence tests for the FLRW, the ``1+log'' GR gauge, and radiation GR gauge are shown in Figs~\ref{fig:flrwtest}, \ref{fig:fractest1} and \ref{fig:fractest2} respectively. We See that generally our simulations are sufficiently convergent for the lower part of the GW spectrum with the $192^3$ resolution. The conservation of the Hamiltonian constraint is also  kept well under control throughout the whole duration of the simulations, and roughly a 3rd order of convergence is achieved; see Fig.~\ref{fig:constraints}.


\section{GWs in the lattice}
\label{sec:proj}

\begin{figure}
\centering
\includegraphics[scale=0.5]{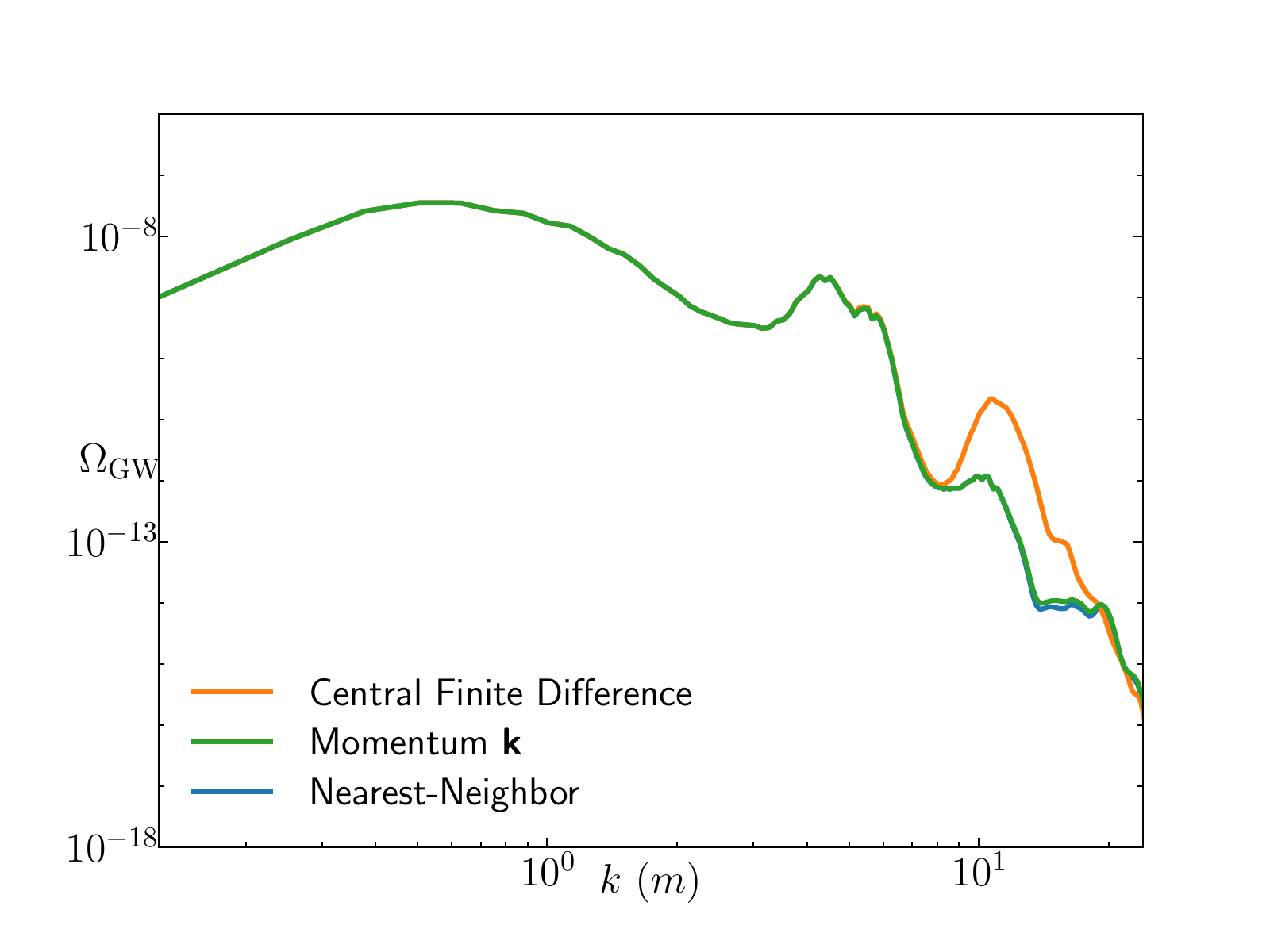}
\caption{GW spectra in the FLRW scheme for Model \RNCaps{3} at $t \approx 172 m^{-1}$ for three projectors: projector based on the central finite difference, conventional projector based on $\textbf{k}$ and projector based on the symmetric nearest-neighbor derivatives.}
\label{fig:projector}
\end{figure}

We now discuss the approach to obtain the GW spectra in the lattice applied to both the FLRW and GR scheme. In order to derive a discrete expression for Eq.~\eqref{eq:gwspectrum} in a lattice of volume $V=L^{3}$, we need to first specify our convention for the discrete Fourier transform: 
\begin{align}
\label{eq:fourier}
f(\mathbf{n}) &=\frac{1}{N^{3}} \sum_{\tilde{n}} e^{-\frac{2 \pi i}{N} \tilde{\mathbf{n}} \mathbf{n}} \tilde{f}(\tilde{\mathbf{n}}) \\
\tilde{f}(\tilde{\mathbf{n}}) &=\sum_{n} e^{+\frac{2 \pi i}{N} \tilde{\mathbf{n}} \mathbf{n}} f(\mathbf{n})
\end{align}
where $\mathbf{n}=\left(n_{1}, n_{2}, n_{3}\right)$ and $\tilde{\mathbf{n}}=\left(\tilde{n}_{1}, \tilde{n}_{2}, \tilde{n}_{3}\right)$ label the lattice sites in configuration space and in the momentum space respectively. It was shown in \cite{Figueroa:2011ye}
\begin{align}
\left(\frac{d \rho_{G W}}{d \log k}\right)(\tilde{\mathbf{n}}) \equiv \frac{d x^{6}k^{3}(|\tilde{\mathbf{n}}|)}{(4 \pi)^{3} G L^{3}}\left\langle\dot{h}_{i j}(|\tilde{\mathbf{n}}|, t)\dot{h}_{i j}^{*}(|\tilde{\mathbf{n}}|, t)\right\rangle_{R(\tilde{\mathbf{n}})}
\end{align}
where $\left\langle\dot{h}_{i j}(|\tilde{\mathbf{n}}|, t) \dot{h}_{i j}^{*}(|\tilde{\mathbf{n}}|, t)\right\rangle_{R(\tilde{\mathbf{n}})}$ is an average over configurations with lattice momenta $\tilde{\mathbf{n}}^{\prime} \in[|\tilde{\mathbf{n}}|,|\tilde{\mathbf{n}}|+\delta \tilde{n}]$. 
As mentioned in the main text, the TT part of tensor perturbation $h_{ij}$ is obtained in Fourier space via the projector $\Lambda_{i j, l m}(\hat{\mathbf{k}})$ given in Eq.~(\ref{eq:lambda}). In the lattice, we want to construct a projector $\Lambda_{i j, l m}$ that satisfies the "transversality" and "tracelessness" conditions, which may come in different forms \cite{Figueroa:2011ye}. 
Nevertheless, for our case, the differences caused by choosing different projectors are very small, less than the percentage level prior to the Nyquist frequency band, as we show in Fig.~\ref{fig:projector}. Thus, we choose the more stable projectors 
\begin{equation}
\begin{aligned}
\Lambda_{i j, l m}^{(L)}(\tilde{\mathbf{n}}) &\equiv P_{i l}^{(L)}(\tilde{\mathbf{n}}) P_{j m}^{(L)}(\tilde{\mathbf{n}})-\frac{1}{2} P_{i j}^{(L)}(\tilde{\mathbf{n}}) P_{l m}^{(L)}(\tilde{\mathbf{n}}) \\
P_{i j}^{(L)}(\tilde{\mathbf{n}}) &=\delta_{i j}-\frac{k_{i}^{(L)} k_{j}^{(L)}}{\left|k^{(L)}\right|^{2}},~~
k_{i}^{(L)}=2 \frac{\sin \left(\pi \tilde{n}_{i} / N\right)}{d x},
\end{aligned}
\end{equation}
which are based on the nearest-neighbour spatial derivatives \cite{Figueroa:2011ye} and fit well with our finite difference scheme in our code.

\bibliography{inspire_auto,manual}

\end{document}